\documentclass[12pt, draftclsnofoot, journal, letterpaper, onecolumn]{IEEEtran}

\usepackage{amsfonts}
\usepackage[dvips]{graphicx}
\usepackage{times}
\usepackage{cite}
\usepackage{amsmath}
\usepackage{cases}
\usepackage{array}
\usepackage{dsfont}
\usepackage{amssymb}

\usepackage{stfloats}
\usepackage{slashbox}
\usepackage{graphicx}
\usepackage{subfigure}
\usepackage{footnote}
\usepackage{booktabs}
\usepackage{array}
\usepackage{bm}
\usepackage{float}
\usepackage{color}
\usepackage{epstopdf}
\newtheorem{remark}{Remark}

\begin{document}

\title{Degrees of Freedom for MIMO Two-Way X Relay Channel}

\author{\IEEEauthorblockN{Zhengzheng Xiang, Meixia Tao, \IEEEmembership{Senior Member, IEEE}, Jianhua Mo, and Xiaodong Wang, \IEEEmembership{Fellow, IEEE}}
\thanks{Copyright (c) 2012 IEEE. Personal use of this material is permitted. However, permission to use this material for any other purposes must be obtained from the IEEE by sending a request to pubs-permissions@ieee.org.}
\thanks{Z. Xiang, M. Tao and J. Mo are with the Department of
Electronic Engineering at Shanghai Jiao Tong University, Shanghai,
200240, P. R. China. Email: \{7222838, mxtao, mjh\}@sjtu.edu.cn.}\thanks{X. Wang is with the Department of Electrical Engineering at Columbia University, New York, USA. Email: wangx@ee.columbia.edu.}
\thanks{This work is supported by the Joint Research Fund for Overseas Chinese,
Hong Kong and Macao Young Scholars under grant 61028001, the Innovation
Program of Shanghai Municipal Education Commission under grant 11ZZ19,
and the NCET program under grant NCET-11-0331.}
}
 \maketitle
\vspace{-1cm}

\begin{abstract}
We study the degrees of freedom (DOF) of a multiple-input multiple-output (MIMO) two-way X relay
channel, where there are two groups of source nodes and one relay node, each equipped with multiple
antennas, and each of the two source nodes in one group exchanges independent messages with the two
source nodes in the other group via the relay node. It is assumed that every source node is equipped with $M$
antennas while the relay is equipped with $N$ antennas. We first show that the upper bound on the total DOF for this network is $2\min\left\{2M,N\right\}$
and then focus on the case of $N\leq 2M$ so that the DOF is upper bounded by twice the number of antennas at the relay. By applying signal alignment for network coding and joint transceiver design for interference cancellation,
we show that this upper bound can be achieved when $N\leq\lfloor\frac{8M}{5}\rfloor$. We also show that with signal
alignment only but no joint transceiver design, the upper bound is achievable when $N\leq\lfloor\frac{4M}{3}\rfloor$.
Simulation results are provided to corroborate the theoretical results and to demonstrate the performance of the proposed
scheme in the finite signal-to-noise ratio regime.

\end{abstract}

\section{Introduction}
Wireless communication has been advancing at an exponential rate,
propelled by the ever-increasing demands for wireless multimedia services.
This, in turn, necessitates the development of novel signaling
techniques with high spectrum efficiency and capacity. Among those
factors limiting the capacity of wireless networks, interference has
been considered as a key bottleneck. Recently, two advanced signaling schemes
have been proposed to cope with interference and to enhance spectrum efficiency: network
coding and interference alignment.

Network coding was originally proposed in \cite{Ahlswede} to achieve the max-flow
bound for the wireline network. The key idea of network coding is to let an intermediate
node combine the messages it receives and forward the mixture to several
destinations simultaneously. Compared with the conventional time-sharing
based schemes where different destinations are served at different
time slots, network coding can increase the overall
throughput significantly. The first wireless application of network coding was the two-way relay channel,
where two source nodes exchange information with the help of
a relay (sometimes referred to as physical layer network coding)
\cite{ZhangSL,Katabi}. By applying physical layer network
coding at the relay, the spectrum efficiency of the two-way relay channel can be doubled compared
with the conventional schemes. Physical layer network coding has also been applied to several
other relay-aided wireless networks such as multiuser two-way relay
networks \cite{WangRui,Sayed,Molisch}, multipair two-way relay channels \cite{Mxtao,Gesbert,Orlik} and
multi-way relay networks \cite{Poor,Kellett,Vucetic}.

Interference alignment was first proposed in
\cite{jafar,Khandani} to achieve the maximum degrees of freedom (DOF) for the
multiple-input multiple-output (MIMO) X channel. It has been shown that
for the MIMO X channel with every node equipped with $M$ antennas,
its total DOF is $\frac{4M}{3}$. The key idea is to align the interference signals so that they
occupy the smallest signal space, leaving more free space for the useful
signals. It was shown in \cite{Jafar1} that the capacity of a $K$-user time-varying interference channel is
characterized by $C({\sf SNR})=\frac{K}{2}\log({\sf SNR})+o(\log({\sf SNR}))$.
Thus, independent of the network size, it is
theoretically possible that each user achieves half
the DOF of an interference-free system. Hence interference is not a fundamental
limitation for such networks. A number of interference alignment schemes
have been proposed, such as distributed interference alignment, ergodic
alignment and blind interference alignment \cite{Jafar2,Jafar3,Jafar4}. An overview on various interference alignment techniques
is given in \cite{Jafar5}.

Based on the concept of interference alignment, signal alignment was proposed in \cite{Chun}
 to solve the network information flow problem for the MIMO Y channel,
where there are three users and a single relay, and each user sends information
to the other two users via the relay. Unlike interference alignment, the goal of signal
alignment is to align the signal streams for different user pairs at the relay. Combined with
network coding, it can significantly increase the network's throughput. In \cite{Namyoon,Ding},
signal alignment was applied to the generalized $K$-user Y channel.

In this paper, we consider the network information flow problem for the MIMO two-way X
relay channel and analyze its total DOF. In this network, there are two groups of
source nodes with each group consisting of two nodes, and a relay node. Each source node in one group
exchanges independent messages with the two source nodes in the other group with the help of the common
relay. It is assumed that every source node is equipped with $M$ antennas and the relay node is equipped
with $N$ antennas. As for practical scenarios of the proposed network information flow,
we can find many applications to wireless networks. For example, in a cooperative multicell
communication system with two base stations and two users connected via a relay, the relay helps exchange
data between the base stations and the users. Also, in a wireless mesh or ad hoc network, two users in one
group exchange information with the two users in the other group via a relay node. We first show that the DOF
of this network is upper bounded by $2\min\left\{2M,N\right\}$. By combining the techniques of signal alignment
for network coding and joint transceiver design for interference cancellation, we then propose an
efficient transmission scheme and show that this scheme achieves the upper bound when $N\leq \left\lfloor\frac{8M}{5}\right\rfloor$.
We also show that with signal alignment only but no joint transceiver design, the upper bound
is achievable when $N\leq \left\lfloor\frac{4M}{3}\right\rfloor$. Note that the MIMO two-way X
relay channel has been considered in \cite{K_Lee} for a special case of $M=3, N=4$. In this paper,
we consider the general case with arbitrary $M$ and $N$. Moreover, when $N\geq \lfloor\frac{4M}{3}\rfloor$ our proposed
scheme outperforms the generalized version of the scheme in \cite{K_Lee}.

The remainder of this paper is organized as follows. In Section II, the
system model of the MIMO two-way X relay channel is described. In Section III, we derive an upper bound
on the DOF of this channel. In Section IV, we present an efficient transmission scheme and give a necessary condition, i.e., $N\leq\left\lfloor\frac{8M}{5}\right\rfloor$, for this scheme
to achieve the upper bound. In Section V, we show that the necessary condition is also sufficient.
In Section VI, we consider a special variate of our proposed transmission scheme
which reduces to the method in \cite{K_Lee} when $M=3, N=4$. Simulation results are provided in Section VII. Finally, Section VIII concludes the paper.

{\sl Notations}: Boldface uppercase letters denote matrices and
boldface lowercase letters denote vectors. $\mathds{R}$, $\mathds{C}$ and $\mathds{Z}^{+}$
denote the sets of real numbers, complex numbers, and positive integers, respectively. $\lfloor x\rfloor=\max\{n\in\mathds{Z}^{+}|n\leq x\}$. $(\cdot)^\textsl{T}$, $(\cdot)^\textsl{H}$, $(\cdot)^\dag$ and $\mbox{Tr}\{\cdot\}$ are the transpose, Hermitian transpose, Moore-Penrose pseudoinverse and
trace operators, respectively. $\mathds{E}(\cdot)$ is the expectation
operator. $\mbox{Span} ({\bf H})$ and ${\mbox{Null} ({\bf H})}$
stand for the column space and the null space of the matrix ${\bf
H}$, respectively. $\mbox{dim}({\bf H})$ denotes the dimension of the column space of ${\bf H}$. ${\bf I}_N$ denotes the $N\times N$
identity matrix and $\oplus$ is the exclusive-OR operator.

\section{MIMO Two-Way X Relay Channel}
\begin{figure}
\begin{centering}
\includegraphics[scale=0.7]{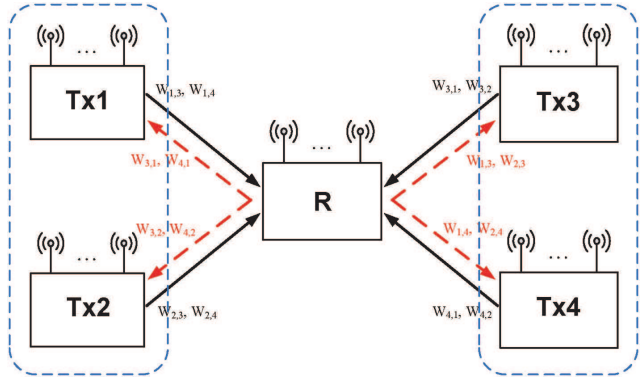}
\vspace{-0.1cm} \caption{MIMO two-way X relay channel.}
\label{fig:Bi_MIMO_X}
\end{centering}
\vspace{-0.3cm}
\end{figure}

Consider a MIMO two-way X relay channel shown in Fig.~\ref{fig:Bi_MIMO_X}. The channel
consists of four source nodes with $M$ antennas each and a relay with $N$ antennas. Each
source node $i$, for $i=1,2$ on the left-hand side (LHS) needs to send an independent
message, denoted as $W_{i,i'}$ to each source node $i'$, for $i'=3,4$ on the right-hand
side (RHS) via the relay. So does each source node on the RHS.

The transmission is implemented in two phases. In the multiple-access (MAC) phase, all four source nodes
transmit their signals to the relay. The received signal at the
relay is given by
\begin{equation}\label{MAC}
{\bf y}_r=\sum\limits_{i=1}^{4}{\bf H}_{i,r}{\bf x}_i +{\bf n}_r
\end{equation}
where ${\bf y}_r$ and ${\bf n}_r$ denote the $N\times 1$ received
signal vector and the additive white Gaussian noise (AWGN) vector at the relay, respectively; ${\bf x}_i$ is the
$M\times 1$ transmitted signal vector by source node $i$ with the power
constraint $\mathds{E}\left(\mbox{Tr}\{{\bf x}_i{\bf
x}_i^H\}\right)\leq P$; ${\bf H}_{i,r}$ is the $N\times M$ channel
matrix from source node $i$ to the relay. The entries of the
channel matrices ${\bf H}_{i,r}$ for, $i=1,2,3,4$, and those of the noise vector ${\bf n}_r$, are independent
and identically distributed (i.i.d.) zero-mean complex Gaussian
random variables with unit variance, i.e., $\mathcal{CN}(0,1)$.
Hence, all channel matrices are of full rank with probability
$1$.

After receiving the signals from the source nodes, the relay forms a new signal ${\bf x}_r$ and broadcasts it to all source
nodes, which is known as the broadcast (BC) phase. The received
signal at the $i$th source node is given by
\begin{equation}\label{BC}
{\bf y}_i={\bf H}_{r,i}{\bf x}_r+{\bf n}_i,~i=1,2,3,4
\end{equation}
where ${\bf y}_i$ and ${\bf n}_i$ denote the $M\times 1$ received
signal vector and the AWGN vector at the
$i$th source node, respectively; ${\bf x}_r$ is the $N\times 1$
transmitted signal vector by the relay with the power constraint
$\mathds{E}\left(\mbox{Tr}\{{\bf x}_r{\bf x}_r^H\}\right)\leq P$;
${\bf H}_{r,i}$ is the $M\times N$ channel matrix from the relay to
source node $i$. Similar to the MAC phase, we assume that ${\bf H}_{r,i}$
and ${\bf n}_i$ contain i.i.d. $\mathcal{CN}(0,1)$ random variables.

Throughout this paper, it is assumed that perfect channel state information (CSI)
is available at all source nodes and the relay\footnote{In practical systems, the required channel state information can be obtained by the relay and the source nodes through the limited feedback techniques, and these part
of information can be transmitted through the backhaul link.}.
Additionally, we assume that the source nodes and the relay operate in full-duplex mode.

We define the total DOF of the above MIMO two-way relay X channel as
\begin{eqnarray}\label{sum_DOF}
\nonumber ~d\triangleq d_{1,3}+d_{1,4}+d_{2,3}+d_{2,4}+d_{3,1}+d_{3,2}+d_{4,1}+d_{4,2}\\
 =\lim\limits_{{\sf SNR}\rightarrow \infty}\frac{R\left({\sf SNR}\right)}{\log\left({\sf SNR}\right)}~~~~~~~~~~~~~~~~~~~~~~~~~~~~~~~~~~~~~
\end{eqnarray}
where $d_{i,j}$ is the DOF from source node $i$ to source node $j$, and $R\left({\sf SNR}\right)$ is the
sum rate as a function of SNR, where SNR is defined as ${\sf SNR}\triangleq P$ since the noise samples are assumed to have unit variance.

\section{An Upper Bound on DOF}
In this section, we derive an upper bound on the DOF of the MIMO two-way X relay channel.

\begin{figure}
\begin{centering}
\includegraphics[scale=0.7]{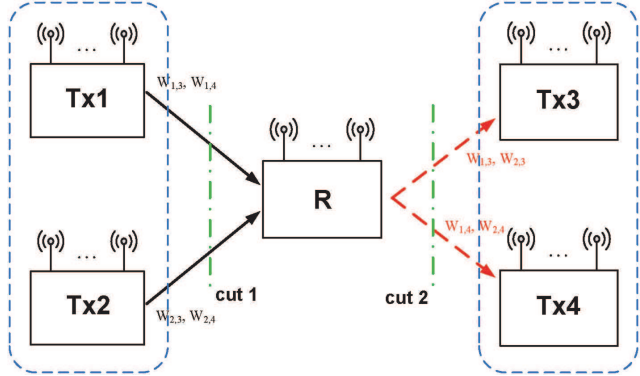}
\vspace{-0.1cm} \caption{One direction of network information flow for
the MIMO two-way X relay channel.}
\label{fig:Bi_MIMO_X_one_direction}
\end{centering}
\vspace{-0.3cm}
\end{figure}

{\emph{\textbf{Theorem} 1: }} Consider a MIMO two-way X relay channel with $M$ antennas at every source node and $N$
antennas at the relay. The total number of DOF is upper bounded by $2\min\left\{2M,N\right\}$, i.e.,
\begin{equation}\label{th1}
d\leq 2\min\left\{2M,N\right\}.
\end{equation}
\begin{proof}
We first consider the network information flow of one direction, i.e., from source nodes $1,2$ to source nodes $3,4$ via the relay, as shown in Fig.~\ref{fig:Bi_MIMO_X_one_direction}.

In the MAC phase (cut $1$), source nodes $1,2$ simultaneously transmit information to the relay.
Assuming source nodes $1$ and $2$ fully cooperate, the channel essentially becomes a $2M\times N$ MIMO channel, whose DOF is $\min\left\{2M,N\right\}$ \cite{Zheng}. In the BC phase (cut $2$), we can obtain the similar result. Applying the cut-set theorem \cite{Cover} to each phase with regard to the DOF, we can have
\begin{eqnarray}\nonumber
d_{1,3}+d_{1,4}+d_{2,3}+d_{2,4}~~~~~~~~~~~~~~~~~~~~~~~~~~~~~~~~~~~~\\\label{D1}\leq \min\big\{\min\left\{2M,N\right\},\min\left\{2M,N\right\}\big\}=\min\left\{2M,N\right\}
\end{eqnarray}

For the other direction of the network information flow, we can similarly obtain
\begin{eqnarray}\label{D2}
d_{3,1}+d_{3,2}+d_{4,1}+d_{4,2}\leq \min\left\{2M,N\right\}.
\end{eqnarray}
Combining \eqref{D1}, \eqref{D2} and using the definition in \eqref{sum_DOF}, we conclude \eqref{th1} which completes the proof.
\end{proof}

\begin{remark}\label{r1}
The factor $2$ on the RHS of \eqref{th1} is due to the assumption of full-duplex mode in our scheme. The same assumption is also used in \cite{Chun}. If half-duplex mode is assumed, the factor of $2$ is not needed.
\end{remark}

From the above result, we can see that when $N\leq 2M$, the total
DOF for the MIMO two-way X relay channel is upper bounded by twice the
number of antennas at the relay, which is therefore the bottleneck for the spectrum efficiency
of the network. In the remainder of the paper, we assume that $N\leq 2M$ so that the upper bound
on the DOF is $2N$. Since the transmission scheme for the case of $N>2M$ will be completely different
from that for the case of $N\leq 2M$, we will leave the case of $N>2M$ to future work.

\section{Efficient Transmission Scheme}
For relay-aided bidirectional channels, applying physical layer
network coding at the relay can significantly improve the system's
spectrum efficiency; and for multiuser channels, beamforming is typically employed to
nulled out the multiuser interference. In this section, by applying signal alignment for network
coding and joint transceiver design for interference cancellation, we propose a novel transmission
scheme, named as ``Signal Alignment with Joint Interference Cancellation (SAJIC)" for the MIMO two-way
relay X channel to maximize its total DOF.

\subsection{A Motivating Example for $M=5,N=8$}

As an example, we consider a system where each
source node has $M=5$ antennas and the relay has $N=8$ antennas. For this system,
the proposed transmission scheme achieves $d_{1,3}=d_{1,4}=d_{2,3}=d_{2,4}=d_{3,1}=d_{3,2}=d_{4,1}=d_{4,2}=2$.
In particular, source node $1$ transmits codewords
$s_{1,3}^1,s_{1,3}^2$ ($s_{1,4}^1,s_{1,4}^2$) for message $W_{1,3}$ ($W_{1,4}$) by using
beamforming vectors ${\bf v}_{1,3}^1,{\bf v}_{1,3}^2$ (${\bf v}_{1,4}^1,{\bf v}_{1,4}^2$), respectively to source node
$3$ (source node $4$) via the relay. Similarly for the other three source nodes.
\begin{figure}
\begin{centering}
\includegraphics[scale=0.7]{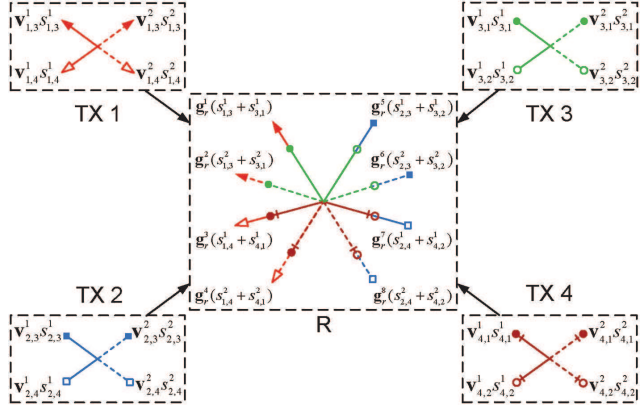}
\vspace{-0.1cm} \caption{Signal alignment for network coding during
the MAC phase.} \label{fig:Bi_MIMO_X_sg}
\end{centering}
\vspace{-0.3cm}
\end{figure}

{\bf{\emph{Step 1:}}} {\bf Signal alignment during the MAC phase}

During the MAC phase, there are totally $16$ data streams arriving at
the relay. Since the relay has only $8$ antennas, it is impossible
for it to decode all the $16$ data streams. However, based on the
idea of physical layer network coding, the relay node only needs
to decode some mixtures of the symbols. Specifically the key point of the
proposed scheme is to obtain the network coded messages
$W_{1,3}\oplus W_{3,1},W_{1,4}\oplus W_{4,1},W_{2,3}\oplus W_{3,2}$
and $W_{2,4}\oplus W_{4,2}$ at the relay (Note that each message consists of two streams.). Inspired by the signal
alignment for network coding \cite{Chun}, we design
the beamformers so that the two desired signals for network coding are
aligned within the same spatial dimension. Taking source node $1$ as
an example, we align its transmitted data streams with the
streams from source node $3,4$ as follows
\begin{eqnarray}\nonumber
&\mbox{span}\left({\bf H}_{1,r}{\bf
v}_{1,3}^1\right)=\mbox{span}\left({\bf H}_{3,r}{\bf
v}_{3,1}^1\right)\triangleq{\bf g}_r^1\\\nonumber&\mbox{span}\left({\bf
H}_{1,r}{\bf v}_{1,3}^2\right)=\mbox{span}\left({\bf H}_{3,r}{\bf
v}_{3,1}^2\right)\triangleq{\bf g}_r^2\\\nonumber&\mbox{span}\left({\bf H}_{1,r}{\bf
v}_{1,4}^1\right)=\mbox{span}\left({\bf H}_{4,r}{\bf
v}_{4,1}^1\right)\triangleq{\bf g}_r^3\\&\mbox{span}\left({\bf
H}_{1,r}{\bf v}_{1,4}^2\right)=\mbox{span}\left({\bf H}_{4,r}{\bf
v}_{4,1}^2\right)\triangleq{\bf g}_r^4
\end{eqnarray}
where ${\bf g}_r^1,{\bf g}_r^2,{\bf g}_r^3,{\bf g}_r^4$ are the signal vectors seen by the
relay. Fig.~\ref{fig:Bi_MIMO_X_sg} illustrates the notion of the signal alignment in the
MAC phase where it is seen that there are $8$ network coded symbols aligned along $8$ signal
vectors, respectively. With $N=8$ antennas, the relay can then obtain the above $8$ network coded symbols.

{\bf{\emph{Step 2:}}} {\bf Joint transceiver design for interference cancellation during
the BC phase}

\begin{figure}
\begin{centering}
\includegraphics[scale=0.7]{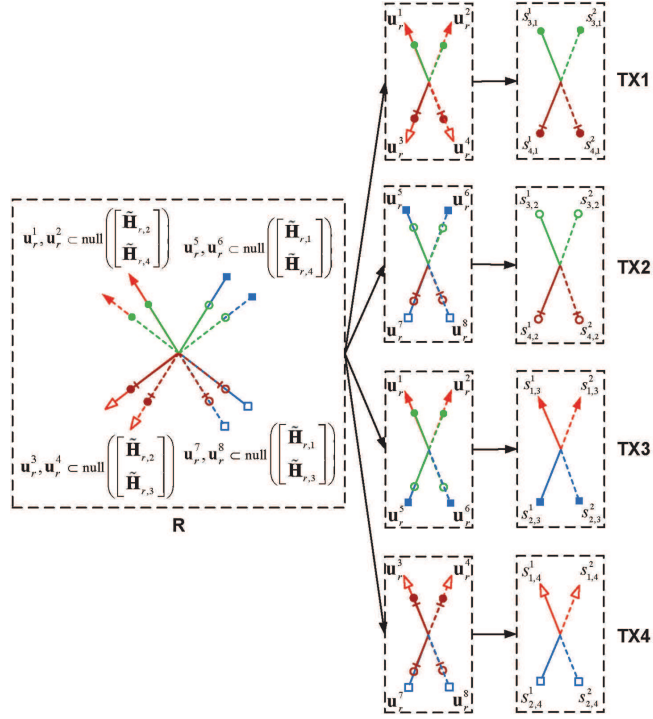}
\vspace{-0.1cm} \caption{Joint interference cancellation during the
BC phase.} \label{fig:Bi_MIMO_X_JIC}
\end{centering}
\vspace{-0.3cm}
\end{figure}

During the BC phase, the relay broadcasts these four network coded
messages using beamformers ${\bf u}_r^1,...,{\bf u}_r^8$. More specifically,
$[{\bf u}_r^1,{\bf u}_r^2],~[{\bf u}_r^3,{\bf u}_r^4],~[{\bf u}_r^5,{\bf u}_r^6],~[{\bf u}_r^7,{\bf u}_r^8]$
are for messages $W_{1,3}\oplus W_{3,1},W_{1,4}\oplus W_{4,1},W_{2,3}\oplus W_{3,2},W_{2,4}\oplus W_{4,2}$
, respectively. Note that at the receiver side each source node suffers from two sources of interference and each component of the
transmitted signal causes interference to two source nodes. For instance, source node $1$ suffers from the interference
caused by $[{\bf u}_r^5,{\bf u}_r^6],~[{\bf u}_r^7,{\bf u}_r^8]$; and $[{\bf u}_r^1,{\bf u}_r^2]$ causes interference to
source nodes $2,4$. Each source node employs a linear receiver, with the $4\times 5$ receiving filter matrix denoted as
${\bf D}_i$, for $i=1,2,3,4$. Denote the effective channel matrix from the relay to source node $i$ as
$\tilde{{\bf H}}_{r,i}\triangleq {\bf D}_i{\bf H}_{r,i}$. The goal of transmit beamformer design at the relay is to make
each component of the transmitted signal to lie in the null space of the effective channel matrices of those unintended source nodes.
For example, $[{\bf u}_r^1,{\bf u}_r^2]$ should satisfy the following condition
\begin{eqnarray}\label{zf1}
[{\bf u}_r^1,{\bf u}_r^2]\subseteq \mbox{Null}\left(\left[\begin{array}{c}\tilde{{\bf H}}_{r,2}\\\tilde{{\bf H}}_{r,4}\end{array}\right]\right)
=\mbox{Null}\left(\left[\begin{array}{c}{\bf d}_{2,3}^1{\bf H}_{r,2}\\{\bf d}_{2,3}^2{\bf H}_{r,2}\\{\bf d}_{2,4}^1{\bf H}_{r,2}\\{\bf d}_{2,4}^2{\bf H}_{r,2}\\
{\bf d}_{4,1}^1{\bf H}_{r,4}\\{\bf d}_{4,1}^2{\bf H}_{r,4}\\{\bf d}_{4,2}^1{\bf H}_{r,4}\\{\bf d}_{4,2}^2{\bf H}_{r,4}
\end{array}\right]\right)
\end{eqnarray}
where ${\bf d}_{i,j}^k$ is the $1\times M$ receiving filter vector for source node $i$ to extract the $k$th data stream from
source node $j$.

However, the dimension of the matrix on the RHS of \eqref{zf1} is $8\times8$ and in general it is full-rank
if each source node chooses its receiving vectors $\{{\bf d}_{i,j}^k\}$ independently. Therefore it is not
possible to find $[{\bf u}_r^1,{\bf u}_r^2]$ that satisfies \eqref{zf1}. In order to circumvent this problem, we
consider joint transceiver design for the source nodes and the relay. More specifically, for source nodes $2,4$, we constrain their receiving vectors to satisfy
\begin{eqnarray}\nonumber
\mbox{span}\left(\left({\bf d}_{2,4}^1{\bf H}_{r,2}\right)^T\right)=\mbox{span}\left(\left({\bf d}_{4,2}^1{\bf H}_{r,4}\right)^T\right)~\\
\mbox{span}\left(\left({\bf d}_{2,4}^2{\bf H}_{r,2}\right)^T\right)=\mbox{span}\left(\left({\bf d}_{4,2}^2{\bf H}_{r,4}\right)^T\right).
\end{eqnarray}
That is, the effective channel matrices for the network coded message $W_{2,4}\oplus W_{4,2}$ from the relay to source nodes $2$ and $4$ are aligned. Then we can choose the transmitting beamformers $[{\bf u}_r^1,{\bf u}_r^2]$ that satisfy
\begin{eqnarray}\label{zf2}
[{\bf u}_r^1,{\bf u}_r^2]\subseteq \mbox{Null}\left(\left[\begin{array}{c}\tilde{{\bf H}}_{r,2}\\\tilde{{\bf H}}_{r,4}\end{array}\right]\right)
=\mbox{Null}\left(\left[\begin{array}{c}{\bf d}_{2,3}^1{\bf H}_{r,2}\\{\bf d}_{2,3}^2{\bf H}_{r,2}\\{\bf d}_{2,4}^1{\bf H}_{r,2}\\{\bf d}_{2,4}^2{\bf H}_{r,2}\\
{\bf d}_{4,1}^1{\bf H}_{r,4}\\{\bf d}_{4,1}^2{\bf H}_{r,4}
\end{array}\right]\right).
\end{eqnarray}
which is feasible since the dimension of the concatenated effective channel matrices is degraded to $6\times8$. The other transmitting and receiving beamformers are designed similarly so that the interference streams are nulled out for each source node.
Fig.~\ref{fig:Bi_MIMO_X_JIC} illustrates the process of joint transceiver design for interference cancellation between source nodes and the relay in the BC phase.

\subsection{Necessary Condition for $d=2N$ when $N\leq2M$}
From the above subsection, we can see that the proposed scheme SAJIC achieves
the DOF upper bound for $M=5,N=8$. In this subsection, we analyze the condition
to achieve the DOF upper bound when $N\leq2M$.

In order to maximize the DOF, it is intuitive
that the number of data streams between each pair of communicating source nodes
should be the same, i.e., $d_{1,3}=d_{3,1}\triangleq d_{13},~d_{1,4}=d_{4,1}\triangleq d_{14},$~$d_{2,3}=d_{3,2}\triangleq d_{23},~d_{2,4}=d_{4,2}\triangleq d_{24}$\footnote{If there is one pair of source nodes which send different number of data streams, the source node with fewer data streams can send extra data streams without increasing the signal space to align with the extra data streams transmitted by the other source node.}. Note that the DOF indicates the maximum number of independent data streams that can be  simultaneously transmitted in the network.

Based on the signal alignment, it can be seen that the dimension of the intersection space of the
channels of each source node pairs determines the maximum number of data streams we can align.
Taking source nodes $1,3$ as an example, according to the {\sl dimension theorem} \cite{Strang} and due to the linear independence of the channel matrcies, we obtain that
\begin{eqnarray}
\nonumber\mbox{dim}\left(\mbox{span}({\bf H}_{1,r})\cap\mbox{span}({\bf
H}_{3,r})\right)~~~~~~~~~~~~~~~~~\\\nonumber=\mbox{dim}\left(\mbox{span}({\bf
H}_{1,r})\right)+\mbox{dim}\left(\mbox{span}({\bf H}_{3,r})\right)~~~~~~~~~\\\nonumber-\mbox{dim}\left(\mbox{span}([{\bf H}_{1,r}~{\bf
H}_{3,r}])\right)~~~~~~~~~~~~~~~~~~~~~~\\\nonumber=\min\left\{M,N\right\}+\min\left\{M,N\right\}-\min\left\{2M,N\right\}
\label{dim2}\\=2\min\left\{M,N\right\}-N.~~~~~~~~~~~~~~~~~~~~~~~~~~~~~~
\end{eqnarray}
which implies that
\begin{eqnarray}\label{dim3}
d_{13}\leq 2\min\left\{M,N\right\}-N.
\end{eqnarray}
For the other pairs of communicating source nodes, we similarly have
\begin{eqnarray}
\label{dim4} \{d_{14}, d_{23}, d_{24}\}\leq 2\min\left\{M,N\right\}-N.
\end{eqnarray}
Combining \eqref{dim3}-\eqref{dim4}, we have
\begin{eqnarray}\label{fea_align}
d_{13}+d_{14}+d_{23}+d_{24}\leq 8\min\left\{M,N\right\}-4N.
\end{eqnarray}
In order to achieve the upper bound, we must satisfy the following condition
\begin{eqnarray}\label{fea_bound}
d_{13}+d_{14}+d_{23}+d_{24}=N.
\end{eqnarray}
Based on \eqref{fea_align} and \eqref{fea_bound}, we obtain
\begin{eqnarray}\nonumber
5N\leq 8\min\left\{M,N\right\}
\end{eqnarray}
which is equivalent to
\begin{eqnarray}
N\leq\left\lfloor\frac{8M}{5}\right\rfloor.
\end{eqnarray}
Thus, we have obtained the necessary condition to achieve the DOF upper bound when
$N\leq 2M$ for the MIMO two-way X relay channel.

\section{Achievability of the Upper Bound}
In this section, we generalize SAJIC in Section IV.A. to
arbitrary $N, M$ with $N\leq2M$, and show that it achieves the DOF upper bound when
$N\leq\left\lfloor\frac{8M}{5}\right\rfloor$. Therefore the necessary condition in Section IV.B
to achieve the DOF upper bound is also sufficient.

We first provide the transmission scheme for the case of $N=\frac{8M}{5}, \forall
M=5k, k\in \mathds{Z}^{+}$ and show that $(d_{1,3}, d_{1,4}, d_{2,3},
d_{2,4}, d_{3,1}, d_{3,2}, d_{4,1}, d_{4,2})=(\frac{N}{4},\frac{N}{4},\frac{N}{4},\frac{N}{4},\frac{N}{4},\frac{N}{4}
,\frac{N}{4},\frac{N}{4})$ is achieved by this scheme.

During the MAC phase, the $i$th source node sends message $W_{i,j}$
to the $j$th source node using $\frac{N}{4}$ independently encoded
streams along beamforming vectors ${\bf V}_{i,j}=\left[{\bf
v}_{i,j}^1,...,{\bf v}_{i,j}^{\frac{N}{4}}\right]$. For instance, the transmitted signal from source node $1$ is
\begin{eqnarray}\nonumber
{\bf x}_1={\bf V}_{1,3}{\bf s}_{1,3}+{\bf V}_{1,4}{\bf
s}_{1,4}~~~~~~\\=\sum\limits_{k=1}^{\frac{N}{4}}\left({\bf
v}_{1,3}^ks_{1,3}^k+{\bf v}_{1,4}^ks_{1,4}^k\right)
\end{eqnarray}
where ${\bf s}_{1,3}$ and ${\bf s}_{1,4}$ are the $\frac{N}{4}\times
1$ encoded symbol vectors for $W_{1,3}$ and $W_{1,4}$, respectively.
The transmitted signals from other source nodes are in a similar
form. In order for the relay to obtain the network coded messages
$W_{1,3}\oplus W_{3,1},W_{1,4}\oplus W_{4,1},W_{2,3}\oplus W_{3,2}$
and $W_{2,4}\oplus W_{4,2}$, we should carefully choose the
beamforming vectors to satisfy the following signal alignment conditions

\begin{eqnarray}\nonumber
{\bf H}_{1,r}{\bf v}_{1,3}^k={\bf H}_{3,r}{\bf
v}_{3,1}^k\triangleq{\bf g}_r^k,~~~~~~~~~~~~~~~~~~~\\\nonumber{\bf
H}_{1,r}{\bf v}_{1,4}^k={\bf H}_{4,r}{\bf v}_{4,1}^k\triangleq{\bf
g}_r^{\frac{N}{4}+k},~~~~~~~~~~~~~~~\\\nonumber{\bf H}_{2,r}{\bf
v}_{2,3}^k={\bf H}_{3,r}{\bf v}_{3,2}^k\triangleq{\bf
g}_r^{\frac{N}{2}+k},~~~~~~~~~~~~~~~\\\label{signal_alignment2}{\bf H}_{2,r}{\bf
v}_{2,4}^k={\bf H}_{4,r}{\bf v}_{4,2}^k\triangleq{\bf
g}_r^{\frac{3N}{4}+k}, ~1\leq k\leq \frac{N}{4}
\end{eqnarray}
where ${\bf g}_r^1,...,{\bf g}_r^N$ are $N$ transmitting vectors
seen by the relay. The above conditions imply that
\begin{eqnarray}\nonumber
\mbox{span}\left(\left[{\bf g}_r^1,...,{\bf
g}_r^{\frac{N}{4}}\right]\right)\subseteq ~\mbox{span}\left({\bf
H}_{1,r}\right)\cap\mbox{span}\left({\bf
H}_{3,r}\right)~~~~~~\\\nonumber\mbox{span}\left(\left[{\bf
g}_r^{\frac{N}{4}+1},...,{\bf
g}_r^{\frac{N}{2}}\right]\right)\subseteq~ \mbox{span}\left({\bf
H}_{1,r}\right)\cap\mbox{span}\left({\bf
H}_{4,r}\right)~~\\\nonumber\mbox{span}\left(\left[{\bf
g}_r^{\frac{N}{2}+1},...,{\bf
g}_r^{\frac{3N}{4}}\right]\right)\subseteq~ \mbox{span}\left({\bf
H}_{2,r}\right)\cap\mbox{span}\left({\bf
H}_{3,r}\right)~\\\mbox{span}\left(\left[{\bf
g}_r^{\frac{3N}{4}+1},...,{\bf g}_r^{N}\right]\right)\subseteq~
\mbox{span}\left({\bf H}_{2,r}\right)\cap\mbox{span}\left({\bf
H}_{4,r}\right).
\end{eqnarray}

Since all the entries of the channel matrices are i.i.d. zero-mean
complex Gaussian random variables, there exists a
$\left(2M-N=\frac{N}{4}\right)$-dimensional intersection subspace
constituted by the column space of channel matrices for each pair of communicating
source nodes with probability $1$. Then we can always choose $\frac{N}{4}$
linearly independent transmitting vectors $\left\{{\bf
g}_r^k\right\}$ for each source node pair. As a result, the received
signal in \eqref{MAC} is rewritten as follows
\begin{eqnarray}
{\bf y}_r={\bf G}_r{\bf s}_r+{\bf n}_r
\end{eqnarray}
where the $N\times N$ matrix ${\bf G}_r\triangleq\left[{\bf
g}_r^1,...,{\bf g}_r^N\right]$, and the $N\times1$ vector
${\bf s}_r\triangleq[s_{1,3}^1+s_{3,1}^1,...,s_{1,3}^{\frac{N}{4}}+s_{3,1}^{\frac{N}{4}},
s_{1,4}^1+s_{4,1}^1,...,s_{1,4}^{\frac{N}{4}}+s_{4,1}^{\frac{N}{4}},
s_{2,3}^1+$ $s_{3,2}^1,...,s_{2,3}^{\frac{N}{4}}+s_{3,2}^{\frac{N}{4}},
s_{2,4}^1+s_{4,2}^1,...,s_{2,4}^{\frac{N}{4}}+s_{4,2}^{\frac{N}{4}}]^T$.
Also since the entries of all channel matrices are independently Gaussian, the probability
that a basis vector in the intersection space of one pair of source
nodes' channel matrices lies in the intersection space of
another pair is zero. Thus ${\bf G}_r$ is full-rank with probability
1, which guarantees the decodability of ${\bf s}_r$ at the relay.
The four network coded messages $\hat{W}_{13}=W_{1,3}\oplus W_{3,1},
\hat{W}_{14}=W_{1,4}\oplus W_{4,1},\hat{W}_{23}=W_{2,3}\oplus
W_{3,2}$ and $\hat{W}_{24}=W_{2,4}\oplus W_{4,2}$ are then obtained
by applying the mapping principle of physical layer network coding \cite{ZhangSL} to each
entry of ${\bf s}_r$.

For the BC phase, the relay broadcasts the network coded messages
$\hat{W}_{13},~\hat{W}_{14},~\hat{W}_{23}$ and $\hat{W}_{24}$ to all
source nodes using encoded symbols ${\bf
q}_r=[q_r^1,...,q_r^{N}]^T$ along the beamforming vectors ${\bf
U}_r=[{\bf u}_r^1,...,{\bf u}_r^N]$. More specifically,
$[q_r^1,...,q_r^{\frac{N}{4}}]^T$,
$[q_r^{\frac{N}{4}+1},...,q_r^{\frac{N}{2}}]^T$,
$[q_r^{\frac{N}{2}+1},...,q_r^{\frac{3N}{4}}]^T$ and
$[q_r^{\frac{3N}{4}+1},...,q_r^{N}]^T$ are the $\frac{N}{4}\times 1$
encoded symbol vectors for
$\hat{W}_{13},~\hat{W}_{14},~\hat{W}_{23}$ and $\hat{W}_{24}$,
respectively. Then the transmitted signal at the relay in \eqref{BC}
is rewritten as
\begin{equation}
{\bf x}_r=\sum\limits_{k=1}^N{\bf u}_r^kq_r^k.
\end{equation}
The received signal at source node $1$ is given by

\begin{eqnarray}\nonumber
\hat{{\bf y}}_1=\tilde{{\bf H}}_{r,1}{\bf x}_r+\tilde{{\bf
n}}_1~~~~~~~~~~~~~~~~~~~~~~~~~~~~~~~~~~~~~~~~~~~~~~~~~~~
\\\nonumber={\bf D}_1{\bf
H}_{r,1}\Bigg(\underbrace{\sum\limits_{k=1}^{\frac{N}{2}}{\bf
u}_r^kq_r^k}_{\mbox{signal}}+\underbrace{\sum\limits_{k=\frac{N}{2}+1}^{\frac{3N}{4}}{\bf
u}_r^kq_r^k+\sum\limits_{k=\frac{3N}{4}+1}^{N}{\bf
u}_r^kq_r^k}_{\mbox{interference}}\Bigg)+\tilde{{\bf n}}_1
\end{eqnarray}
Recall that the matrices $\{\tilde{{\bf H}}_{r,i}\}$ and $\{{\bf D}_{r,i}\}$ are defined in Section IV.A. The first term in the bracket represents the combination of
the desired network-coded messages $\hat{W}_{13}$ and
$\hat{W}_{14}$, while the remaining two terms are the unwanted
interference $\hat{W}_{23}$ and $\hat{W}_{24}$. The received signals
at the other source nodes are written in a similar way.

Next, we jointly design the transceivers for the source nodes and the relay for interference
cancellation. Due to the symmetry of the MAC and BC phases, we can design the receiving matrix on each source node such that the effective receiving channels of each source node pair are aligned:
\begin{eqnarray}\nonumber
&{\bf d}_{1,3}^k{\bf H}_{r,1}={\bf d}_{3,1}^k{\bf H}_{r,3}\triangleq {\bf w}_{13}^k,~~~~~~~~~~~~~~~\\\nonumber
&{\bf d}_{1,4}^k{\bf H}_{r,1}={\bf d}_{4,1}^k{\bf H}_{r,4}\triangleq {\bf w}_{14}^k,~~~~~~~~~~~~~~~\\\nonumber
&{\bf d}_{2,3}^k{\bf H}_{r,2}={\bf d}_{3,2}^k{\bf H}_{r,3}\triangleq {\bf w}_{23}^k,~~~~~~~~~~~~~~~\\\label{receive_align2}
&{\bf d}_{2,4}^k{\bf H}_{r,2}={\bf d}_{4,2}^k{\bf H}_{r,4}\triangleq {\bf w}_{24}^k,~1\leq k\leq\frac{N}{4}.
\end{eqnarray}
Here, ${\bf w}^k_{ij}$ is a $1\times N$ effective channel
vector between source node $i$ and source node $j$ on the $k$-th data stream. Since the signal alignment has been applied successfully in the MAC phase, for the BC phase, each source node can also choose its receiving vectors to satisfy the above conditions, and the resulting $N$ effective channel vectors $\left\{{\bf
w}_{ij}^k\right\}$ are linearly independent with probability $1$.

For the beamforming vectors at the relay, we can choose them to lie in the intersection subspace of each source
node pair's effective channels' null space as follows
\begin{eqnarray}\nonumber
&\mbox{span}\left(\left[{\bf u}_r^1,...,{\bf
u}_r^{\frac{N}{4}}\right]\right)~~~~~~\subseteq
\mbox{Null}\left(\left[\begin{array}{c}\tilde{{\bf H}}_{r,2}\\\tilde{{\bf H}}_{r,4}\end{array}\right]\right)~\\\nonumber&\mbox{span}\left(\left[{\bf
u}_r^{\frac{N}{4}+1},...,{\bf
u}_r^{\frac{N}{2}}\right]\right)~~\subseteq
\mbox{Null}\left(\left[\begin{array}{c}\tilde{{\bf H}}_{r,2}\\\tilde{{\bf H}}_{r,3}\end{array}\right]\right)~\\\nonumber&\mbox{span}\left(\left[{\bf
u}_r^{\frac{N}{2}+1},...,{\bf
u}_r^{\frac{3N}{4}}\right]\right)~\subseteq
\mbox{Null}\left(\left[\begin{array}{c}\tilde{{\bf H}}_{r,1}\\\tilde{{\bf H}}_{r,4}\end{array}\right]\right)~\\\label{IC2}&\mbox{span}\left(\left[{\bf
u}_r^{\frac{3N}{4}+1},...,{\bf u}_r^{N}\right]\right)~\subseteq
\mbox{Null}\left(\left[\begin{array}{c}\tilde{{\bf H}}_{r,1}\\\tilde{{\bf H}}_{r,3}\end{array}\right]\right).
\end{eqnarray}
We show that there exists a $\frac{N}{4}$-dimensional null space for the concatenated effective channel matrix
of each pair of communicating source nodes with probability $1$. Taking source nodes $2$ and $4$ as an example, the
dimension of $\left[\tilde{{\bf H}}_{r,2}^T,~\tilde{{\bf H}}_{r,4}^T\right]^T$ is $N\times N$. Since we have aligned
their receiving effective channels in \eqref{receive_align2}, $\left[\tilde{{\bf H}}_{r,2}^T,~\tilde{{\bf H}}_{r,4}^T\right]^T$
has $\frac{N}{4}$ repeated rows and its rank is $\min\{N-\frac{N}{4},N\}=\frac{3N}{4}$. Therefore, the dimension of its null space is $N-\frac{3N}{4}=\frac{N}{4}$. For the the other source node pairs, we can similarly get
the result.

{\emph{\textbf{Lemma} 1: }} During the BC phase, the null space of the concatenated effective
channel matrix for each source node pair has no intersection with that of
the other source node pairs, i.e.,
\begin{eqnarray}\nonumber
\mbox{Null}\left(\left[\begin{array}{c}\tilde{{\bf H}}_{r,i}\\\tilde{{\bf H}}_{r,j}\end{array}\right]\right)\cap
\mbox{Null}\left(\left[\begin{array}{c}\tilde{{\bf H}}_{r,m}\\\tilde{{\bf H}}_{r,n}\end{array}\right]\right)=\phi,
~\forall (i,j)\neq (m,n).
\end{eqnarray}

\begin{proof}
We first consider the concatenated effective channel matrices for source node pairs $(1,3)$ and $(1,4)$,
and have that
\begin{eqnarray}\nonumber
\mbox{Null}\left(\left[\begin{array}{c}\tilde{{\bf H}}_{r,1}\\\tilde{{\bf H}}_{r,3}\end{array}\right]\right)\cap
\mbox{Null}\left(\left[\begin{array}{c}\tilde{{\bf H}}_{r,1}\\\tilde{{\bf H}}_{r,4}\end{array}\right]\right)~~~~~~~~~~~~~~\\\nonumber=\mbox{Null}
\left(\left[\begin{array}{c}\tilde{{\bf H}}_{r,1}\\\tilde{{\bf H}}_{r,3}\\\tilde{{\bf H}}_{r,4}\end{array}\right]\right)=\mbox{Null}\left(\left[{{\bf w}_{13}^1}^T,...,{{\bf w}_{13}^{{\frac{N}{4}}}}^T,\right.\right.~~~~\\\label{lemma1}\left.\left.{{\bf w}_{14}^1}^T,...,{{\bf w}_{14}^{{\frac{N}{4}}}}^T,{{\bf w}_{23}^1}^T,...,{{\bf w}_{23}^{{\frac{N}{4}}}}^T,{{\bf w}_{24}^1}^T,...,{{\bf w}_{24}^{{\frac{N}{4}}}}^T\right]^T\right)
\end{eqnarray}
Since the $N\times N$ matrix at the end of \eqref{lemma1} is full-rank, the dimension of its null space is always
zero. For the other pair of source nodes, the same argument holds and the lemma follows.
\end{proof}

According to Lemma $1$, it can be seen that all the $N$ beamforming vectors $\left\{{\bf
u}_r^k\right\}$ at the relay are linearly independent with probability $1$. Thus the received signals at
source node $1$ can be rewritten as
\begin{eqnarray}
\hat{{\bf y}}_1=\tilde{{\bf H}}_{r,1}\bigg(\underbrace{\sum\limits_{k=1}^{\frac{N}{4}}{\bf
u}_r^kq_r^k}_{\mbox{for}~
\hat{W}_{13}}+\underbrace{\sum\limits_{k=\frac{N}{4}+1}^{\frac{N}{2}}{\bf
u}_r^kq_r^k}_{\mbox{for}~ \hat{W}_{14}}\bigg)+\tilde{{\bf n}}_1.
\end{eqnarray}
Thus, there is no interference for source node $1$ and it can decode
these useful signals. Then by using its own messages, source node $1$
can obtain the messages from source nodes $3,4$ as follows
\begin{eqnarray}
& W_{3,1}=W_{1,3}\oplus \hat{W}_{13}, ~W_{4,1}=W_{1,4}\oplus
\hat{W}_{14}.
\end{eqnarray}
In the same manner, the other source nodes can also obtain the
messages intended for themselves. Therefore, a total of $2N$ DOF
is achieved by using the proposed scheme on MIMO two-way X relay channel.
\subsection{$N\leq\left\lfloor\frac{8M}{5}\right\rfloor$ or $M \neq 5k$}
For the other cases that $M\neq 5k$ or $N\leq\left\lfloor\frac{8M}{5}\right\rfloor$, we can choose
the DOF for each pair for different values
of $N$ as below
\begin{eqnarray}\nonumber
&N=4k:~~~~~~~~~~~~~~~~~~~~~~~~~~~~~~~~~~~~~~~~~~~~~~~~~~~~~~\\\nonumber&d_{1,3}=d_{3,1}=\frac{N}{4},~d_{1,4}=d_{4,1}=\frac{N}{4}~~~~~~~~~~~~~~
\\\nonumber&d_{2,3}=d_{3,2}=\frac{N}{4},~d_{2,4}=d_{4,2}=\frac{N}{4}~~~~~~~~~~~~~~\\\nonumber
&N=4k+1:~~~~~~~~~~~~~~~~~~~~~~~~~~~~~~~~~~~~~~~~~~~~~~~~~~\\\nonumber&d_{1,3}=d_{3,1}=\left\lfloor\frac{N}{4}\right\rfloor,~d_{1,4}=d_{4,1}=\left\lfloor\frac{N}{4}\right\rfloor
~~~~~~~~\\\nonumber&d_{2,3}=d_{3,2}=\left\lfloor\frac{N}{4}\right\rfloor,~d_{2,4}=d_{4,2}=\left\lfloor\frac{N}{4}\right\rfloor+1~~~~
\\\nonumber&N=4k+2:~~~~~~~~~~~~~~~~~~~~~~~~~~~~~~~~~~~~~~~~~~~~~~~~~~\\\nonumber&d_{1,3}=d_{3,1}=\left\lfloor\frac{N}{4}\right\rfloor,~d_{1,4}=d_{4,1}=\left\lfloor\frac{N}{4}\right\rfloor+1~~~~
\\\nonumber&d_{2,3}=d_{3,2}=\left\lfloor\frac{N}{4}\right\rfloor+1,~d_{2,4}=d_{4,2}=\left\lfloor\frac{N}{4}\right\rfloor~~~~
\\\nonumber&N=4k+3:~~~~~~~~~~~~~~~~~~~~~~~~~~~~~~~~~~~~~~~~~~~~~~~~~~\\\nonumber&d_{1,3}=d_{3,1}=\left\lfloor\frac{N}{4}\right\rfloor,~d_{1,4}=d_{4,1}=\left\lfloor\frac{N}{4}\right\rfloor+1~~~~
\\\nonumber&d_{2,3}=d_{3,2}=\left\lfloor\frac{N}{4}\right\rfloor+1,~d_{2,4}=d_{4,2}=\left\lfloor\frac{N}{4}\right\rfloor+1
\end{eqnarray}
We can similarly apply the previous transmission scheme to achieve the upper DOF bound
$2N$ and the process is briefly described as follows.

For the MAC phase, we show that the signals for each pair of source
nodes can be aligned at the relay: since
\begin{eqnarray}\nonumber
N\leq\left\lfloor\frac{8M}{5}\right\rfloor\leq\frac{8M}{5},
\end{eqnarray}
we can have
\begin{eqnarray}
2M-N\geq\left\{\begin{array}{c}\frac{N}{4},~~\mbox{when}~N=4k\\\left\lfloor\frac{N}{4}\right\rfloor+1,~\mbox{when}~N\neq4k.\end{array}\right.
\end{eqnarray}
For the BC phase, the receiving alignment is also feasible just as for
the MAC phase. For the transmitting beamforming design at the relay,
we show that relay can always choose linearly independent
beamforming vectors for each part of the signals. Without loss of
generality, we take $[{\bf u}_r^1,...,{\bf u}_r^{d_{13}}]$ as an
example, which should satisfy the following condition
\begin{eqnarray}
\mbox{span}\left(\left[{\bf u}_r^1,...,{\bf u}_r^{d_{13}}\right]\right)\subseteq
\mbox{Null}\left(\left[\begin{array}{c}\tilde{{\bf H}}_{r,2}\\\tilde{{\bf H}}_{r,4}\end{array}\right]\right).
\end{eqnarray}
Since
\begin{eqnarray}
\tilde{{\bf H}}_{r,2}=\left[\begin{array}{c}{\bf
w}_{23}^1\\\vdots\\{\bf w}_{23}^{d_{23}}\\{\bf
w}_{24}^1\\\vdots\\{\bf w}_{24}^{d_{24}}\end{array}\right],\tilde{{\bf H}}_{r,4}=\left[\begin{array}{c}{\bf w}_{14}^1\\\vdots\\{\bf
w}_{14}^{d_{14}}\\{\bf w}_{24}^1\\\vdots\\{\bf
w}_{24}^{d_{24}}\end{array}\right],
\end{eqnarray}
we have
\begin{eqnarray}\nonumber
\mbox{dim}\left(\mbox{Null}\left(\left[\begin{array}{c}\tilde{{\bf H}}_{r,2}\\\tilde{{\bf H}}_{r,4}\end{array}\right]\right)\right)\\\nonumber=N-d_{23}-d_{24}-d_{14}~~~~~~\\=d_{13}=\mbox{dim}\left([{\bf u}_r^1,...,{\bf
u}_r^{d_{13}}]\right).
\end{eqnarray}
Finally, we summarize the algorithm for SAJIC in the following chart

\vspace{0.4cm} \hrule \hrule \vspace{0.2cm} \textbf{Outline of SAJIC} \vspace{0.2cm} \hrule
\vspace{0.3cm} ~~
\begin{itemize}
\item {\bf Step 1.}~ In the MAC phase, each source node designs its beamforming vectors $\{{\bf v}_{i,j}^k\}$ according to
\eqref{signal_alignment2} so that the two desired signals for network coding are aligned at the relay node.
\item {\bf Step 2.}~ By applying the mapping principle of physical layer network coding, the relay then decodes its received signals to obtain the
network coded messages $\{\hat{W}_{ij}\}$.
\item {\bf Step 3.}~ In the BC phase, the source nodes and the relay jointly design their transceivers. More specifically, all the source nodes design
their receiving filter matrices $\{{\bf D}_i\}$ according to \eqref{receive_align2}; the relay designs its transmitting beamforming vectors $\{{\bf u}_r^k\}$ according to \eqref{IC2} to cancel the interference for each source node.
\item {\bf Step 4.}~ Each source node decodes its received signals to obtain the network coded messages $\{\hat{W}_{ij}\}$ intended for itself. Using its side-information, each node finally acquires its desired messages $\{W_{i,j}\}$.
\end{itemize}
\vspace{0.2cm} \hrule \vspace{0.4cm}

\begin{remark}\label{r2}
We can see that the DOF for each
source node may not be the same when $N$ is not a multiple of $4$.
However, we can apply four time slots extension here to let every source
node achieve the same DOF $\frac{N}{2}$. Specifically, when using time extension of $4$ channel uses,
the channel is equivalent to a $4M\times4N$ MIMO two-way X relay channel in which each source node has $4M$ antennas and the relay has $4N$ antennas. Then our proposed scheme SAJIC can be applied to this situation directly and each node achieves the equal DOF of $\frac{N}{2}$.
\end{remark}

In Section IV.B we have shown that $N\leq\left\lfloor\frac{8M}{5}\right\rfloor$ is a necessary condition to achieve the DOF upper bound.
And in this section we have shown that this condition is also sufficient. Hence we have the following main result of this paper.

{\emph{\textbf{Theorem} 2: }} When $N\leq2M$, the necessary and sufficient condition for SAJIC to achieve the
DOF upper bound $2N$ in the MIMO two-way X relay channel is $N\leq\left\lfloor\frac{8M}{5}\right\rfloor$.

\begin{remark}
When $N\leq M$, the DOF upper bound $2N$ can also be achieved by applying standard techniques in
two-way relay channel. More specifically, we only allow one source node pair to use the relay in one time slot and apply time sharing among different source node pairs. In each time slot, the network just reduces to the standard two-way relay channel
and those existing techniques can then be used. It can be seen that this simple method can achieve $2\min\left\{N,M\right\}$. However, it can no longer achieve the DOF upper bound when $N>M$.
\end{remark}

Next, we analyze the achievable DOF when $N\leq\left\lfloor\frac{8M}{5}\right\rfloor$ is not satisfied. When $\left\lfloor\frac{8M}{5}\right\rfloor<N<2M$, although our proposed scheme SAJIC cannot achieve the DOF upper bound $2N$, it can still work. More specifically, the dimension of intersection space for each source node pair's channel matrices is $2M-N$. Following the outline in previous sections, it can be seen that the total DOF SAJIC can achieve is $8\cdot(2M-N)=16M-8N$. Compared with the DOF upper bound, the gap is $2N-(16M-8N)=10N-16M$. When $N\geq 2M$, SAJIC is not feasible since the dimension of intersection space for each source node pair's channel matrices is zero.

\section{Connection with the Transmission Method in \cite{K_Lee}}

In the previous sections, we have shown that using our SAJIC, the cut-set outer
bound for the DOF can be achieved when $N\leq\left\lfloor\frac{8M}{5}\right\rfloor$.
Specifically, we align the signals for each pair of source nodes in the MAC phase and apply joint
transceiver design for interference cancellation in the BC phase. In this section, we will show that if we do not
consider the joint transceiver design but directly apply interference nulling beamforming at the relay in the
BC phase, our proposed scheme will reduce to a generalized version of the transmission method in \cite{K_Lee}.

For the reduced or simplified transmission scheme which does not apply the joint transceiver design in the BC phase,
we consider as an example the case $N=\frac{4M}{3}, \forall M=3k, k\in \mathds{Z}^{+}, (d_{1,3}, d_{1,4}, d_{2,3}, d_{2,4},$
$d_{3,1}, d_{3,2}, d_{4,1}, d_{4,2})=(\frac{N}{4},\frac{N}{4},\frac{N}{4},\frac{N}{4},\frac{N}{4},\frac{N}{4} ,\frac{N}{4},\frac{N}{4})$.
In the MAC phase, we similarly apply signal alignment as in \eqref{signal_alignment2}. In the
BC phase, since we directly apply interference nulling at the relay, the relay
will cancel one part of interference for each source node, leaving the remaining part of interference to be cancelled by
the source node itself. More specifically, the relay can choose its beamformers as
\begin{eqnarray}\nonumber
\mbox{span}\left(\left[{\bf u}_r^1,...,{\bf
u}_r^{\frac{N}{4}}\right]\right)~~~~~~\subseteq \mbox{Null}({\bf
H}_{r,4})~\\\nonumber\mbox{span}\left(\left[{\bf
u}_r^{\frac{N}{4}+1},...,{\bf
u}_r^{\frac{N}{2}}\right]\right)~~\subseteq \mbox{Null}({\bf
H}_{r,2})~\\\nonumber\mbox{span}\left(\left[{\bf
u}_r^{\frac{N}{2}+1},...,{\bf
u}_r^{\frac{3N}{4}}\right]\right)~\subseteq \mbox{Null}({\bf
H}_{r,1})~\\\mbox{span}\left(\left[{\bf
u}_r^{\frac{3N}{4}+1},...,{\bf u}_r^{N}\right]\right)~\subseteq
\mbox{Null}({\bf H}_{r,3}).
\end{eqnarray}
For each channel matrix ${\bf H}_{r,i},i=1,...,4$, there exists a
$\left(N-M=\frac{N}{4}\right)$-dimensional null space with
probability 1. Then for each network coded message, the relay can
choose $\frac{N}{4}$ linearly independent vectors. Also it can be easily seen that
the $N$ beamforming vectors are linearly
independent with probability $1$. Thus the received signals at source node $1$ can be
rewritten as
\begin{eqnarray}
{\bf y}_1={\bf
H}_{r,1}\bigg(\underbrace{\sum\limits_{k=1}^{\frac{N}{2}}{\bf
u}_r^kq_r^k}_{\mbox{signal}}+\underbrace{\sum\limits_{k=\frac{3N}{4}+1}^{N}{\bf
u}_r^kq_r^k}_{\mbox{interference}}\bigg)+{\bf n}_1.
\end{eqnarray}
Note that source node $1$ has $M={\frac{3N}{4}}$ antennas and the dimension of the
useful signal is $\frac{N}{2}$. So it has exactly
$\left(\frac{3N}{4}-\frac{N}{2}=\frac{N}{4}\right)$ free
dimensions for the interference signal whose dimension is also
$\frac{N}{4}$. Thus source node $1$ can cancel the other part of the
interference by itself. More specifically, source node $1$ can choose its receiving
matrix ${\bf D}_1\in \mathds{C}^{\frac{N}{2}\times M}$ as follows
\begin{eqnarray}
\mbox{span}\left({\bf D}_1^T\right)\subseteq \mbox{Null}\left(\left[{\bf H}_{r,1}{\bf u}_r^{\frac{3N}{4}},...,
{\bf H}_{r,1}{\bf u}_r^{N}\right]^T\right).
\end{eqnarray}
Since $\left[{\bf H}_{r,1}{\bf u}_r^{\frac{3N}{4}},...,
{\bf H}_{r,1}{\bf u}_r^{N}\right]$ is an $M\times \frac{N}{4}$ matrix, the dimension of its left null space
is $M-\frac{N}{4}=\frac{N}{2}$. Source node $1$ can choose $\frac{N}{2}$ linearly independent receiving
filter row vectors, and hence ${\bf D}_i$ is full-rank with probability $1$. Then the received signals
for source node $1$ is
\begin{eqnarray}
\hat{{\bf y}}_1={\bf D}_1\bigg(\underbrace{\sum\limits_{k=1}^{\frac{N}{4}}{\bf
u}_r^kq_r^k}_{\mbox{for}~
\hat{W}_{13}}+\underbrace{\sum\limits_{k=\frac{N}{4}+1}^{\frac{N}{2}}{\bf
u}_r^kq_r^k}_{\mbox{for}~ \hat{W}_{14}}\bigg)+\tilde{{\bf n}}_1.
\end{eqnarray}
There is no interference and source node $1$ achieves the DOF of $\frac{N}{2}$. The other source nodes operate in
the same manner. Therefore, the reduced transmission scheme also achieves the total DOF of $2N$.

\begin{remark}
The method given in \cite{K_Lee} corresponds to the reduced transmission scheme for $N=4, M=3$.
\end{remark}

Next, we show that the reduced transmission scheme requires a stricter condition to achieve the DOF upper bound, i.e. $N\leq\left\lfloor\frac{4M}{3}\right\rfloor$.

In the MAC phase, the condition that the reduced scheme needs to satisfy is
\begin{eqnarray}\label{fea_mac1}
N\leq\left\lfloor\frac{8M}{5}\right\rfloor
\end{eqnarray}
which is the same as the original proposed transmission scheme. Extra conditions are needed in the BC phase for the reduced scheme.

{\emph{\textbf{Lemma} 2: }} For the reduced transmission scheme in the BC phase, for each source node, the dimension of
the interference that needs to be canceled by the relay is\footnote{If $N\leq M$, the relay does not need to cancel the interference
and each source node can null all the interference it suffers by itself.} $N-M$.
\begin{proof}
Without loss of generality, we take source node $1$ as an example.
For source node $1$, the dimension of useful signals is $d_{13}+d_{14}$;
while the dimension of interference is $d_{23}+d_{24}$. Since it has
$M$ antennas, the interference dimension that it can
cancel by itself is $M-(d_{13}+d_{14})$. Then the interference dimension that needs to
be nulled at the relay is
\begin{eqnarray}\nonumber
d_{23}+d_{24}-\left[M-(d_{13}+d_{14})\right]\\\nonumber
=d_{13}+d_{14}+d_{23}+d_{24}-M~~\\\nonumber=N-M.~~~~~~~~~~~~~~~~~~~~~~~~
\end{eqnarray}
For the other source nodes, we can similarly obtain the result and
the lemma follows.
\end{proof}

For source node $1$, suppose $d_{23}^1$ out of $d_{23}$ interference
streams and $d_{24}^1$ out of $d_{24}$ interference streams are nulled out
at the relay. Then according to Lemma $2$, we have
\begin{eqnarray}\label{fea_bc1}
d_{23}^1+d_{24}^1=N-M.
\end{eqnarray}
Since the dimension of the null space of ${\bf H}_{r,1}$ is also
$N-M$, the relay can choose beamformers which lie in its channel
matrix's null space to cancel these interference streams. For source node $2$, we
can similarly have that
\begin{eqnarray}\label{fea_bc2}
d_{13}^1+d_{14}^1=N-M.
\end{eqnarray}

We now consider source node $3$ and source node $4$. As for source node $3$, the interference signals consist of $d_{14}+d_{24}$ data streams. From the previous discussion, we know that $d_{14}^1$ out
of $d_{14}$ data streams lie in the null space of source node
$2$'s channel matrix; and $d_{24}^1$ out of $d_{24}$ data streams lie in the null space of source node $1$'s channel matrix. These interference signals cannot lie in the null space of ${\bf H}_{r,3}$ and therefore cannot be nulled out by the relay: the proof technique is similar to that applied in Lemma $1$. Thus, source node $3$ must cancel them by itself, which implies that
\begin{eqnarray}\label{fea_bc3}
d_{14}^1+d_{24}^1\leq M-\left(d_{13}+d_{23}\right).
\end{eqnarray}
Similarly for source node $4$, we should have
\begin{eqnarray}\label{fea_bc4}
d_{13}^1+d_{23}^1\leq M-\left(d_{14}+d_{24}\right).
\end{eqnarray}
Combining \eqref{fea_bc3} and \eqref{fea_bc4}, we have
\begin{eqnarray}\label{fea_bc5}
d_{13}^1+d_{14}^1+d_{23}^1+d_{24}^1\leq 2M-\left(d_{13}+d_{14}+d_{23}+d_{24}\right).
\end{eqnarray}
Plugging \eqref{fea_bound}, \eqref{fea_bc1} and \eqref{fea_bc2} into \eqref{fea_bc5}, we then obtain
\begin{eqnarray}\label{fea_bc6}
N\leq\left\lfloor\frac{4M}{3}\right\rfloor
\end{eqnarray}
which is the condition the reduced scheme should satisfy in the BC phase.
Now combining \eqref{fea_mac1} and \eqref{fea_bc6}, the necessary condition for the reduced
scheme to achieve the upper bound becomes simply \eqref{fea_bc6}.

From the above analysis, it can be seen that using the reduced transmission scheme, we can
achieve the DOF upper bound $2N$ in a range $0<N\leq\left\lfloor\frac{4M}{3}\right\rfloor$. By further applying
joint transceiver design for interference cancellation in the BC phase, our proposed scheme can
achieve the upper bound $2N$ in a wider range $0<N\leq\left\lfloor\frac{8M}{5}\right\rfloor$.

\section{Simulation Results}
In this section, we provide numerical results to show the ergodic sum rate performance for
the proposed transmission scheme. Then, we will demonstrate that the proposed
scheme exactly attains the upper bound on the DOF derived in Section V. The channel is assumed as the normalized
Rayleigh fading channel, i.e., the elements of each channel vector are independent and
identically distributed circularly symmetric zero-mean complex Gaussian random variables with
unit variance. The numerical results are illustrated with respect to the ratio of the total transmitted signal
power to the noise variance at each receive antenna in decibels $(\mbox{SNR}=P)$. Each result is averaged over $10000$ independent channel realizations.

We now explain how we compute the sum rate for the MIMO two-way X relay channel when applying the SAJIC. In the MAC phase, assuming that the zero forcing detector ${\bf F}_r=\left[{{\bf f}_r^1}^T,...,{{\bf f}_r^N}^T\right]^T$ is
applied by the relay, the achievable rate for network coded message $\hat{W}_{13}$ is calculated as
\begin{eqnarray}\nonumber
R_{13}=~~~~~~~~~~~~~~~~~~~~~~~~~~~~~~~~~~~~~~~~~~~~~~~~~~~~~~~~~~~~~~~~~\\\nonumber\log\left[\det\big({\bf I}+[{{\bf w}_r^1}^T,...,{{\bf w}_r^{d_{13}}}^T]^T{\bf G}_r{\bf G}_r^H[{{\bf w}_r^1}^H,...,{{\bf w}_r^{d_{13}}}^H]\big)\right]~~~~
\end{eqnarray}
In the BC phase, the achievable rate for $\hat{W}_{13}$ at source node $1$ and $3$ is given by
\begin{eqnarray}\nonumber
R_{13}'=~~~~~~~~~~~~~~~~~~~~~~~~~~~~~~~~~~~~~~~~~~~~~~~~~~~~~~~~~~~~~~~\\\nonumber\log\left[\det\big({\bf I}+[{{\bf w}_{13}^1}^T,...,{{\bf w}_{13}^{d_{13}}}^T]^T{\bf U}_r{\bf U}_r^H[{{\bf w}_{13}^1}^H,...,{{\bf w}_{13}^{d_{13}}}^H]\big)\right]
\end{eqnarray}
Then we have
\begin{eqnarray}
R_{1,3}=R_{3,1}=\min\left\{R_{13},~R_{13}'\right\}.
\end{eqnarray}
The rates of other pairs of communicating source nodes can be computed in a similar way. Thus, we can obtain the achievable sum rate for the whole network when SAJIC is applied.

\begin{figure}
\begin{centering}
\includegraphics[scale=0.45]{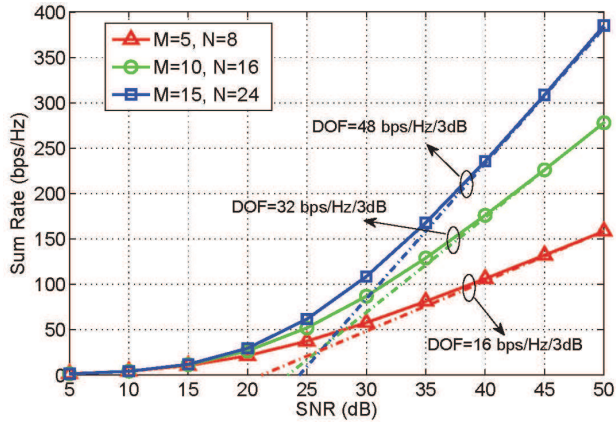}
\vspace{-0.1cm} \caption{The ergodic sum rate for the MIMO two-way X relay channel under different network architectures.}
\label{fig:DOF_M_N}
\end{centering}
\vspace{-0.3cm}
\end{figure}

\begin{figure}
\begin{centering}
\includegraphics[scale=0.45]{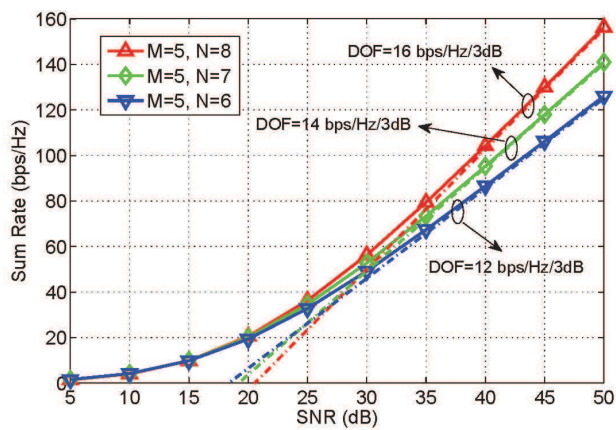}
\vspace{-0.1cm} \caption{The ergodic sum rate for the MIMO two-way X relay channel when $M$ is fixed.}
\label{fig:DOF_5_N}
\end{centering}
\vspace{-0.3cm}
\end{figure}

In Fig.~\ref{fig:DOF_M_N}, we plot the sum rate performance of the proposed scheme according to various antenna configurations. We can see that, as analyzed in Section V, the proposed scheme indeed achieves the upper bound on the DOF. Specifically, we can always observe a sum-rate increase of
$2N$ bps/Hz for every $3$ dB increase in SNR. For instance, when $M = 5,N = 8$, the curve has a slope of $2N=16$. In Fig.~\ref{fig:DOF_5_N}, we plot the sum rate performance of the network when the number of antennas at each source $M$ is fixed. It can be seen that as the number of antennas at the relay $N$ increases, the total DOF also increase, which shows that the relay antenna number is a bottleneck of the network when $N\leq 2M$.

\section{Conclusion}
This paper  considered the total DOF for the
 MIMO two-way X relay channel. We analyzed the upper
bound on the DOF for such a network.
Then by exploiting physical layer network coding and joint
interference cancellation, we proposed SAJIC and
showed that SAJIC can achieve the upper bound if and only if $N\leq\left\lfloor\frac{8M}{5}\right\rfloor$.
Also, we generalized the scheme in \cite{K_Lee} and derived a necessary condition for it to
achieve the upper bound. Besides, we analyzed the relationship between
these two schemes and showed that our proposed SAJIC can be reduced to the
generalized version of the scheme in \cite{K_Lee} after some simplification.

The achievability of the upper bound on the DOF for the considered network in the case of $\left\lfloor\frac{8M}{5}\right\rfloor < N < 2M$
and the case of $N\geq2M$ remains open for further investigation.

\begin{IEEEbiography}{Zhengzheng Xiang}
received the B.S. degree in electronic engineering from Shanghai Jiao Tong University, Shanghai, China, in 2010. He is currently working toward the Ph.D. degree with the Institute of Wireless Communication Technology, Shanghai Jiao Tong University.

His research interests include interference management in wireless networks, wireless relay technologies, and advanced signal processing for wireless cooperative communication.
\end{IEEEbiography}

\begin{IEEEbiography}{Meixia Tao}
(S'00-M'04-SM'10) received the B.S. degree in electronic engineering from Fudan University, Shanghai, China, in 1999, and the Ph.D. degree in electrical and electronic engineering from Hong Kong University of Science and Technology in 2003. She is currently an Associate Professor with the Department of Electronic Engineering, Shanghai Jiao Tong University, China. From August 2003 to August 2004, she was a Member of Professional Staff at Hong Kong Applied Science and Technology Research Institute Co. Ltd. From August 2004 to December 2007, she was with the Department of Electrical and Computer Engineering, National University of Singapore, as an Assistant Professor. Her current research interests include cooperative transmission, physical layer network coding, resource allocation of OFDM networks, and MIMO techniques.

Dr. Tao is an Editor for the \textsc{IEEE Transactions on Communications} and the \textsc{IEEE Wireless Communications Letters}. She was on the Editorial Board of the \textsc{IEEE Transactions on Wireless Communications} from 2007 to 2011 and the \textsc{IEEE Communications Letters} from 2009 to 2012. She also served as Guest Editor for \textsc{IEEE Communications Magazine} with feature topic on LTE-Advanced and 4G Wireless Communications in 2012, and Guest Editor for \textsc{EURISAP J WCN} with special issue on Physical Layer Network Coding for Wireless Cooperative Networks in 2010. She was in the Technical Program Committee for various conferences, including IEEE INFOCOM, IEEE GLOBECOM, IEEE ICC, IEEE WCNC, and IEEE VTC.

Dr. Tao is the recipient of the IEEE ComSoC Asia-Pacific Outstanding Young Researcher Award in 2009 and the co-recipient of the International Conference on Wireless Communications and Signal Processing (WCSP) Best Paper Award in 2012.
\end{IEEEbiography}

\begin{IEEEbiography}{Jianhua Mo}
received the B.S. degree in electronic engineering in Shanghai Jiao Tong University, China, in 2010. He is pursuing a dual M.S. degree from Shanghai Jiao Tong University and Georgia Institute of Technology. His research interest is physical layer security.
\end{IEEEbiography}

\begin{IEEEbiography}{Xiaodong Wang}
(S'98-M'98-SM'04-F'08) received the Ph.D degree in Electrical Engineering from Princeton
University.

He is a Professor of  Electrical Engineering at Columbia University in New York.
Dr. Wang's research interests fall in the general areas of computing, signal processing
and communications, and has published extensively in these areas. Among his
publications is a book entitled ``Wireless Communication Systems: Advanced Techniques for Signal
Reception'', published by Prentice Hall in 2003.  His current research interests include wireless
communications, statistical signal processing, and genomic signal processing.

Dr. Wang received the 1999 NSF CAREER Award, the 2001
IEEE Communications Society and Information Theory Society Joint Paper Award,
and the 2011 IEEE Communication Society Award for Outstanding Paper on New Communication
Topics. He has served  as an Associate Editor for the {\em IEEE Transactions on Communications},
the {\em IEEE Transactions on Wireless Communications}, the {\em IEEE Transactions on Signal Processing},
and the {\em IEEE Transactions on Information Theory}. He is a Fellow of the IEEE and listed as
an ISI Highly-cited Author.
\end{IEEEbiography}


\begin{thebibliography}{99}
\bibitem{Ahlswede} R. Ahlswede, N. Cai, S. R. Li and R. W. Yeung, ``Network
information flow," {\sl IEEE Transactions on Information Theory}, vol.
46, pp. 1204¨C1217, Jul. 2000.


\bibitem{ZhangSL} S. Zhang, S. Liew and P. Lam, ``Physical layer network
coding," {\sl Proc. ACM MobiCom}, pp. 63-68, Sep. 2006.

\bibitem{Katabi} S. Katti, S. Gollakota and D. Katabi, ``Embracing wireless
interference: Analog network coding," {\sl Proc. ACM SIGCOMM}, pp.
397-408, Sep. 2007.

\bibitem{WangRui} R. Wang and M. Tao, ``Linear precoding designs for amplify-and-forward
multiuser two-way relay systems", {\sl Proc. IEEE GLOBECOM}, Houston Texas, USA, 5-9 Dec. 2011.

\bibitem{Sayed} J. Joung and Ali H. Sayed, ``Multiuser two-way
amplify-and-forward relay processing and power control methods for
beamforming systems," {\sl IEEE Transactions on Signal Processing},
vol. 58, no. 3, pp. 1833-1846, Mar 2010.

\bibitem{Molisch} C. Wang, H. Chen, Q. Yin, A. Feng and A. F. Molisch, ``Multi-user
two-way relay networks with distributed beamforming," {\sl IEEE
Transactions on Wireless Communications}, vol. 10, no. 10, pp.
3460-3471, Oct. 2011.

\bibitem{Mxtao} M. Tao and R. Wang, ``Linear precoding for multi-pair two-way MIMO
relay systems with max-min fairness", to appear in {\sl IEEE Transactions on Signal Processing}, 2012.

\bibitem{Gesbert} E. Yilmaz, R. Zakhour, D. Gesbert and R. Knopp, ``Multi-pair
two-way relay channel with multiple antenna relay station," {\sl
Proc. IEEE ICC}, 2010.


\bibitem{Orlik} T. Koike-Akino, M. Pun and P. Orlik, ``Network-coded
interference alignment in K-Pair bidirectional relaying channels,"
{\sl Proc. IEEE ICC}, 2011.

\bibitem{Poor} D. Gundiiz, A. Yener, A. Goldsmith and H. Poor, ``The multi-way relay channel,"
{\sl Proc. IEEE ISIT 2009}, Seoul, Korea, June 28-July 3, 2009.

\bibitem{Kellett} L. Ong, S. Johnson, and C. Kellett, ``An optimal coding strategy for the binary
multi-way relay channel," {\sl IEEE Communication Letters}, vol. 14, no. 4, Apr. 2010.

\bibitem{Vucetic} Z. Zhou, K. Teav and B. Vucetic, ``Beamforming optimization and power allocation for
MIMO asymmetric multi-way relay channels," {\sl IEEE Communication Letters}, vol. 16, no. 6, Jun. 2012.


\bibitem{jafar} S. A. Jafar and S. Shamai (Shitz), ``Degrees of freedom region of the MIMO X channel,"
 {\sl IEEE Transactions on Information Theory}, vol. 54, no. 1, Jan. 2008.

\bibitem{Khandani} M. A. Maddah-Ali, A. S. Motahari, and A. K. Khandani, ``Communication
over MIMO X channels: Interference alignment, decomposition, and
performance analysis," {\sl IEEE Transactions on Information
Theory}, vol. 54, no. 8, Aug. 2008.

\bibitem{Jafar1} V. R. Cadambe and S. A. Jafar, ``Interference alignment and
the degrees of freedom for the K User interference channel," {\sl IEEE
Transactions on Information Theory}, vol. 54, no. 8, Aug. 2008.

\bibitem{Jafar2} K. S. Gomadam, V. R. Cadambe and S. A. Jafar, ``A distributed numerical approach
 to interference alignment and applications to wireless interference networks," {\sl IEEE Transactions
 on Information Theory}, vol. 57, no. 6, Jun. 2011.

\bibitem{Jafar3} S. A. Jafar, ``The ergodic capacity of phase-fading interference networks," {\sl IEEE
Transactions on Information Theory}, vol. 57, no. 12, Dec. 2011.

\bibitem{Jafar4} C. Wang, T. Gou and S. A. Jafar, ``Aiming perfectly in the
dark - blind interference alignment through staggered antenna
switching," {\sl IEEE Transactions on Signal Processing}, vol. 59,
no. 6, Jun. 2011.

\bibitem{Jafar5} S. A. Jafar, ``Interference alignment: a new look at signal dimensions in a communication network,"
{\sl Foundations and Trends in Communications and Information
Theory}, vol. 7, no. 1, 2011.

\bibitem{Chun} N. Lee, J. Lee and J. Chun, ``Degrees of freedom on the MIMO Y
channel: signal space alignment for network coding," {\sl IEEE
Transactions on Information Theory}, vol. 56, no. 7, Jul. 2010.

\bibitem{Namyoon} K. Lee, N. Lee and I. Lee, ``Achievable Degrees of Freedom on
K-user Y Channels," {\sl IEEE Transactions on Wireless Communications}, vol. 11, no. 3,
Mar. 2012.

\bibitem{Ding} N. Wang, Z. Ding, X. Dai and A. Vasilakos, ``On Generalized MIMO Y Channels:
Precoding Design, Mapping, and Diversity Gain," {\sl IEEE Transactions on Vehicular Technology},
vol. 60, no. 7, Sep. 2011.


\bibitem{K_Lee} K. Lee, S.H. Park, J.S. Kim and I. Lee, ``Degrees of Freedom
on MIMO Multi-Link Two-Way Relay Channels," in {\sl Proc. IEEE
Globecom}, December 2010.

\bibitem{Zheng} L. Zheng and D. N. C. Tse, ``Diversity and multiplexing: a fundamental
tradeoff in multiple antenna channels," {\sl IEEE Transactions on Information Theory},
vol. 49, May 2003.

\bibitem{Cover} T. Cover and A. E. Gamal, ``Capacity theorems for the relay
channel," {\sl IEEE Transactions on Information Theory}, vol. 25, no.
5, Sep. 1979.

\bibitem{Strang} G. Strang, ``Linear Algebra and Its Applications", Brooks/Cole; International ed,
4th edition, 2004.

\end{thebibliography}
\end{document}